\documentclass[12pt]{article}
\usepackage{graphicx}
\begin{document}
\title{Simple procedure for classical signal-procession in cluster-state quantum computation}
\author{Kazuto Oshima\thanks{E-mail: oshima@elc.gunma-ct.ac.jp}    \\ \\
\sl Gunma National College of Technology, Maebashi 371-8530, Japan}
\date{}
\maketitle
\begin{abstract}
We exhibit a simple procedure to find how classical signals should be processed 
in cluster-state quantum  computation. Using stabilizers characterizing 
a cluster state, we can easily find a precise classical signal-flow that is required
in performing cluster-state computation.
\end{abstract} 
\newpage
Cluster-state quantum computation proposed by Raussendorf and Briegel\cite{Rausendorf-Briegel}
is a promising scheme for quantum computation. Preparing a cluster-state\cite{Briegel} 
we can perform quantum computation only by successive quantum measurements and
feed forward of measurement outcomes\cite{Rausendorf}.  
In cluster state quantum computation it is inevitable to properly
choose signs of measurement angles and rectify the resultant
quantum state according to preceding random measurement outcomes.
Fundamental gates such as the controlled-not gate,
an arbitrary single-qubit unitary gate and Hadamard gate have already been studied
and given as packages\cite{Browne}. Combining these packages,  an arbitrary computation
is carried out. In each package rather complicated eigenvalue equations originating from 
a cluster-state by successive measurements prescribe how random measurement outcomes
affect the sings of measurement angles and an output state.  
A classical signal-flow in cluster-state computation has also been studied 
in literature\cite{Cashefi1, Cashefi2, Cashefi3, Beaudrap1, Beaudrap2}.

Cluster state quantum computation can be simulated by quantum teleportation
circuit\cite{Nielsen,Gott,KLM}. This scheme will give us more flexibility
in programming. In this scheme more concise programming
and saving the number of qubits will be possible. For example
the original package for a controlled-not gate is made of a cluster state with
15 qubits.  In contrast with this, in the quantum teleportation sheme this operation
can be located on a cluster state as a controlled-phase transformation at the starting point.
In the quantum teleportation scheme a correction according to each random measurement
outcome is required in the course of computation.  
In a practical cluster-state quantum computation these corrections should be pushed
forward to output qubits. By the presence of controlled-phase 
transformations, however, it is not easy to tract effect of these corrections.
Our purpose in this paper is to give a simple procedure to find a precise classical
signal-flow in cluster-state quantum computation with the help of the 
quantum teleportation scheme.  

In a quantum teleportation circuit if all measurements results are 0, we can 
obtain a correct result without any output corrections or any measurement angle corrections.
Therefore if all measurements results are trivial we can perform all controlled-phase transformations
before any measurements and we can operate the quantum teleportation circuit only 
by measuring the corresponding cluster state.
As for general measurement outcomes, 
we can easily see that a measurement result 1 is compensated by an operation
$X$ before the meter. Moreover this operation $X$ is compensated by an operation $Z$
on the corresponding qubit in the cluster state.  Therefore we can see all effects of 
the measurement outcomes by the random $Z$ operations on the cluster.
We show an example of quantum teleportation circuits and its substitute in Fig.1. 
We show that these $Z$ operations can be removed by successive use of stabilizers
\cite{stabilizer1,stabilizer2,stabilizer3,stabilizer4} 
characterizing the cluster state.   

We consider a cluster state of the following type\cite{Briegel}
\begin{equation}
 |\phi\rangle = \Pi_{i} (|0\rangle +|1\rangle {\vec Z})_{i}, 
\end{equation}
where the index $i$ runs over all qubits in the cluster;
we assume that this index gives an order for qubits in the
cluster and the arrow over the Pauli matrix $Z$ means this operation
acts only on adjacent qubits with a number larger than $i$.  Here and in the following
we neglect an overall factor of a state.
This state is characterized by the following stabilizer equations
\begin{equation}
 K_{i}|\phi\rangle =|\phi\rangle, 
\end{equation}
where $K_{i}=X_{i}\Pi_{j}Z_{j}$ with the index $j$
runs over all adjacent qubits of $i$-th qubit. 

It is convenient to divide qubits in a cluster into three parts; input qubits, body
qubits and output qubits. 
Strictly speaking, we do not have a cluster state
even at the starting point, because in general input qubits are not in the
form $(|0\rangle +|1\rangle {\vec Z})_{i}$.  The state $|\Phi\rangle$ we confront
is obtained by replacing the state
$(|0\rangle +|1\rangle {\vec Z})_{i}$ in $|\phi\rangle$ by an arbitrary state  
$(a|0\rangle +b|1\rangle {\vec Z})_{i}$ for input qubits in the cluster. 
This state still satisfies the
stabilizer equations $K_{i}|\Phi\rangle =|\Phi\rangle$ for $i$-th qubit not being
an input qubit.

As stated before, on account of the randomness of the measurement outcomes, the state  $|\Phi\rangle$
changes into the state $\Pi_{i}Z_{i}^{s_{i}}|\Phi\rangle$, where the product is for 
all input qubits and body qubits and $s_{i}$ is a measurement result 0(1) of the $i$-th
qubit.  Since the operator $K_{i+1}$  has the factor $Z_{i}$, operating  $K_{i+1}^{s_{i}}$ on
the state $|\Phi\rangle$, we can remove the $Z_{i}^{s_{i}}$ factor on $|\Phi\rangle$.
Thus, operating the stabilizers $K_{i}$ on $|\Phi\rangle $ adequately, we can remove 
all of the $Z_{i}^{s_{i}}$ factors on $|\Phi\rangle$.  After these operations we
have the following identity
\begin{equation}
 \Pi_{i}Z_{i}^{s_{i}}|\Phi\rangle
=\Pi_{j}X_{j}^{f_{j}}Z_{j}^{g_{j}}\Pi_{k}X_{k}^{h_{k}}|\Phi\rangle,
\end{equation}
where the index $i$ on the left hand side runs over the input and the body qubits as before
and the indices $j$ and $k$ on the right hand side run over the output qubits and the body
qubits, respectively.  The factor $X_{j}^{f_{j}}Z_{j}^{g_{j}}$ is a correction on 
the $j$-th output qubit depending
on the measurement outcomes. Pushing forward the factor $X_{k}^{h_{k}}$ on the
$k$-th body qubit to the measurement meter,
it changes into a harmless operation $Z_{k}^{h_{k}}$. Since 
$X^{s}Z_{\alpha}=Z_{(-1)^{s}\alpha}X^{s}$, after the above procedure 
the measurement angle $\alpha_{k}$ is replaced by 
${(-1)^{h_{k}}\alpha_{k}}$ in the $k$-th qubit. 
In this way we can easily find the output correction factors 
$X_{j}^{f_{j}}Z_{j}^{g_{j}}$ and the signs ${(-1)^{h_{k}}}$ of measurement angles systematically. 

As an example let us consider an arbitrary single-qubit unitary transformation
$U_{z}(\gamma)U_{x}(\beta)U_{z}(\alpha)$. 
This transformation is carried out by five-qubits cluster state in Fig.2\cite{Browne}.
The first qubit is an input qubit and the fifth qubit is an output qubit.
The measurement angles of the first four qubits are $0,\pm\alpha,\pm\beta, \pm\gamma$, respectively.
The $\pm$ factors should be chosen properly depending on the preceding measurement results.
We can easily obtain the following identity
\begin{eqnarray}
 Z_{4}^{s_{4}}Z_{3}^{s_{3}}Z_{2}^{s_{2}}Z_{1}^{s_{1}}|\Phi\rangle 
&=&Z_{4}^{s_{4}}Z_{3}^{s_{3}}Z_{2}^{s_{2}}Z_{1}^{s_{1}}K_{5}^{s_{4}}K_{4}^{s_{3}}
K_{3}^{s_{2}}K_{2}^{s_{1}}|\Phi\rangle   \nonumber\\
&=&X_{5}^{s_{2}+s_{4}}Z_{5}^{s_{1}+s_{3}}X_{4}^{s_{1}+s_{3}}X_{3}^{s_{2}}X_{2}^{s_{1}}
|\Phi\rangle ,
\end{eqnarray}
where we have applied the stabilizer $K_{5}^{s_{4}}K_{4}^{s_{3}}K_{3}^{s_{2}}K_{2}^{s_{1}}$ to remove the
factor $Z_{4}^{s_{4}}Z_{3}^{s_{3}}Z_{2}^{s_{2}}Z_{1}^{s_{1}}$.
From this identity we find the correction factor of the output qubit 
$X_{5}^{s_{2}+s_{4}}Z_{5}^{s_{1}+s_{3}}$ and the measurement angles are fixed
as $(-1)^{s_{1}}\alpha,(-1)^{s_{2}}\beta, (-1)^{s_{1}+s_{3}}\gamma$.

As for a one-dimensional chain we have seen that it is easy to find a correction factor 
and to fix the sign factors.  We also find that the above procedure 
can be easily applied to a cluster state containing controlled transformations.   
Let us consider an H-branch in Fig.3 that contains essential elements for a controlled
transformation.  We assume that the 1-st qubit and 4-th qubit are input qubits and
3-rd qubit and 6-th qubit are output qubits.
Then we have the following state 
$Z_{5}^{s_{5}}Z_{4}^{s_{4}}Z_{2}^{s_{2}}Z_{1}^{s_{1}}|\Phi\rangle$.
To remove the factors $Z_{1}^{s_{1}},Z_{4}^{s_{4}}$ 
we first operate the stabilizer $K_{2}^{s_{1}}K_{5}^{s_{4}}$ on
$|\Phi\rangle$, where $K_{2}=X_{2}Z_{1}Z_{3}Z_{5}$ and  $K_{5}=X_{5}Z_{2}Z_{4}Z_{6}$.
After this operation we have 
$Z_{6}^{s_{4}}Z_{5}^{s_{1}+s_{5}}Z_{3}^{s_{1}}Z_{2}^{s_{2}+s_{4}}X_{5}^{s_{4}}X_{2}^{s_{1}}|\Phi\rangle$.
Next to remove the factors $Z_{2}^{s_{2}+s_{4}},Z_{5}^{s_{1}+s_{5}}$
we operate $K_{3}^{s_{2}+s_{4}}K_{6}^{s_{1}+s_{5}}$ on $|\Phi\rangle$, where
$K_{3}=X_{3}Z_{2}$ and $K_{6}=X_{6}Z_{5}$. We finally have the state
\begin{equation}
Z_{6}^{s_{4}}X_{6}^{s_{1}+s_{5}}Z_{3}^{s_{1}}X_{3}^{s_{2}+s_{4}}X_{5}^{s_{4}}X_{2}^{s_{1}}
|\Phi\rangle.
\end{equation}
From this expression we see the output correction is 
$Z_{6}^{s_{4}}X_{6}^{s_{1}+s_{5}}Z_{3}^{s_{1}}X_{3}^{s_{2}+s_{4}}$
and the 2-nd and 5-th qubits' measurement angles are multiplied by the 
factors $(-1)^{s_{1}}$ and $(-1)^{s_{4}}$. This procedure will work
for more complicated cluster state composed of many H-branches.
 
We have given a procedure to process classical signals in the cluster-state
quantum computation.   We have shown that it is useful to replace random 
measurement outcomes by random $Z^{s}$ operators.  Applying the stabilizers 
to remove the $Z^{s}$ operators, we easily find how classical signals should be
processed. Executing this procedure, it will be not so hard to
compute out a classical signal-flow even for a much more complicated cluster.

\newpage

FIGURE CAPTIONS\\
FIG~1. (a) Single-qubit two-stage quantum teleportation circuit. The meter means
a quantum measurement by the base $\{|0\rangle, |1\rangle\}$. We have two 
random measurement outcomes $s_{1},s_{2}$.\\
(b) Corresponding circuit with each random outcome is replaced by the random $Z^{s}$ operator.
In compensation all measurement results are assumed to be 0.\\ 
\\ \\ \\
FIG~2. (a) One-dimensional cluster state for an arbitrary 
single-qubit unitary transformation. \\
(b) Corresponding circuit with each random outcome is replaced by the random $Z^{s}$ operator.
These $Z^{s}$ operators can be removed by suitable stabilizers.
\\ 
\\ \\ \\
FIG~3. An H-branch that simulates a controlled transformation.\\

\newpage

\begin{figure}[htbp]
\includegraphics[width=0.8\linewidth]{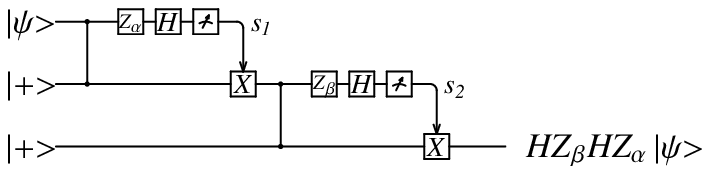} \\
\end{figure}
\vspace{1cm}
FIG.1(a)

\begin{figure}[htbp]
\includegraphics[width=0.8\linewidth]{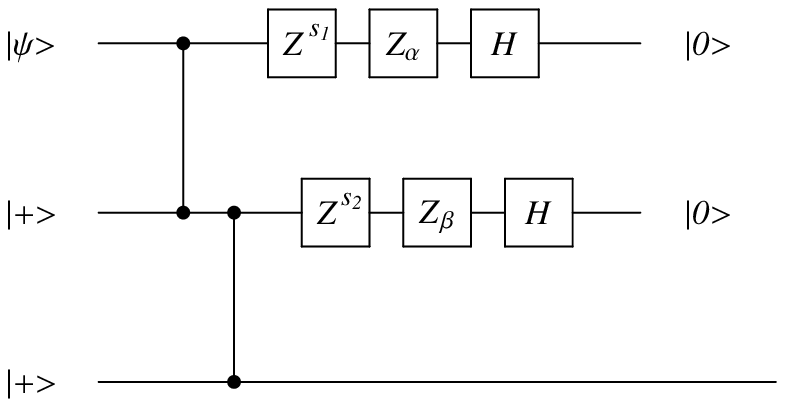} \\
\end{figure}
\vspace{1cm}
FIG.1(b)

\newpage

\begin{figure}[htbp]
\includegraphics[width=0.6\linewidth]{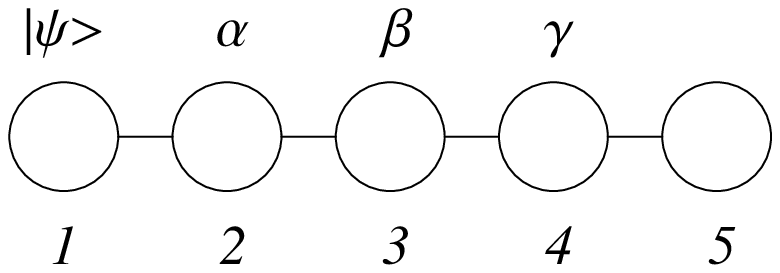}\\
\end{figure}
\vspace{-1cm}
FIG.2(a)\\ \\ \\ 
\begin{figure}[htbp]
\includegraphics[width=0.3\linewidth]{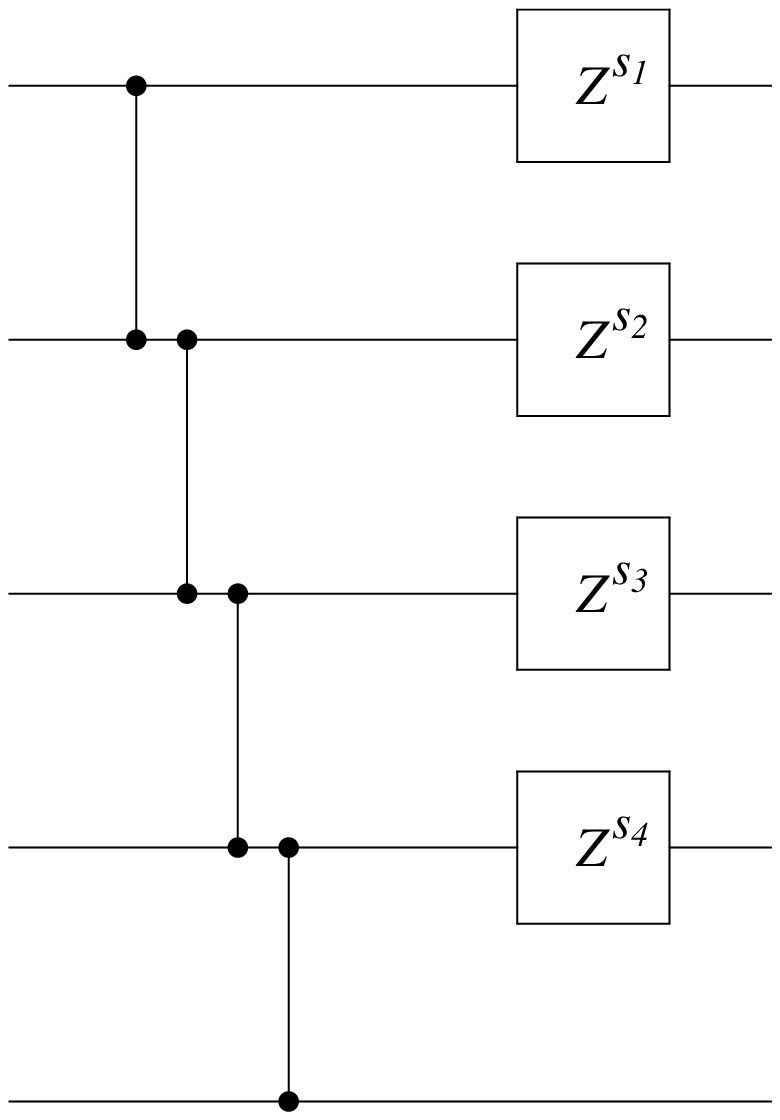}\\
\end{figure}
\\
FIG.2(b)

\newpage

\begin{figure}[htbp]
\includegraphics[width=0.4\linewidth]{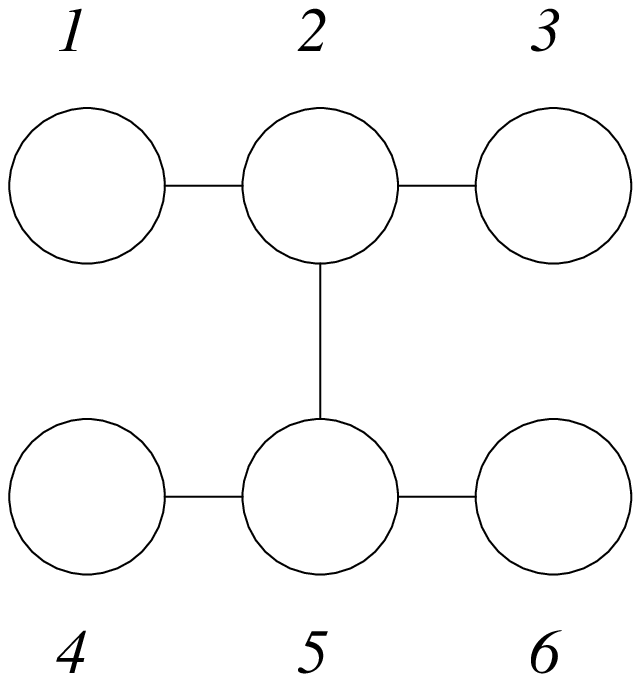}\\
\end{figure}
FIG.3

\end{document}